\documentclass[a4paper,10pt, bibtotoc]{scrartcl}
\pdfoutput=1 
\usepackage{hyperref}\hypersetup{colorlinks=false,linkcolor=blue,citecolor=blue,urlcolor=blue}

\usepackage[T1]{fontenc}
\usepackage[latin1]{inputenc}
\usepackage{fancyhdr}
\usepackage[dvips]{graphics}
\usepackage[dvips]{graphicx}
\usepackage{color}
\usepackage{longtable}
\usepackage{supertabular}
\usepackage{lscape}
\usepackage{afterpage}
\usepackage{setspace}  % Veraenderung des Zeilenabstands
\usepackage{calc}
\usepackage{verbatim}
\usepackage{latexsym}
\usepackage{float,rotating}

\definecolor{leichtgrau}{gray}{.50}
\definecolor{farbe}{rgb}{1,0.05,0.55}
\usepackage{framed, color}
\definecolor{shadecolor}{gray}{.8}

\usepackage{ae}
\usepackage[english]{babel}
\usepackage{subfig}
\usepackage{textcomp}

\usepackage{epsfig}

\usepackage{cancel}

\usepackage{wrapfig}
\usepackage{psfrag}

\usepackage{amsmath}
\usepackage{eufrak}
\usepackage[nointegrals]{wasysym}
\usepackage{footnpag}

\usepackage[numbers,super,sort]{natbib}

\usepackage{remreset}
\usepackage{subfig}
\makeatletter
\@removefromreset{equation}{section}
\renewcommand*\theequation{\@arabic\c@equation}
\@removefromreset{figure}{section}
\@removefromreset{table}{section}

\renewcommand*\thetable{\@arabic\c@table}
\makeatother

\setkomafont{sectioning}{\rmfamily \bfseries}
\setkomafont{captionlabel}{\rmfamily \bfseries}

\usepackage{multicol}
\usepackage{nicefrac}
\usepackage{units}

\usepackage{authblk}

\title{Effective potentials between gold nano crystals - functional dependence on the temperature}
\author[1]{G. Bauer \thanks{bauer@itt.uni-stuttgart.de}}
\author[1]{A. Lange}
\author[1]{N. Gribova}
\author[2]{C. Holm}
\author[1]{J. Groß}
\affil[1]{Institute for Thermodynamics and Thermal Process Engineering, University of Stuttgart}
\affil[2]{Institute for Computational Physics, University of Stuttgart}
\date{}
\begin{document}
\maketitle
\renewcommand{\theequation}{\thesection.\arabic{equation}}
 \begin{abstract}
 
A method is presented that allows to combine the effective potential between two nano crystals, the potential of mean force (PMF), as obtained from all-atomistic Molecular Dynamics simulations with perturbation theory. In this way, a functional dependence of the PMF on temperature is derived, that enables the prediction of the PMF in a wide temperature range. We applied the method for systems of capped gold nano crystals of different size. They show very good agreement with data from atomistic simulations.

\end{abstract}

%%%%%%%%%%%%INTRODUCTION%%%%%%%%%%%% 
\section{Introduction}

Nano crystals (NC) are building blocks of newly engineered materials that combine optical and electrical properties in a custom made fashion.\cite{wangj, west, Shevchenko} Depending on their shape, the solvent, and the temperature, nano crystals can form superstructures, each of which is associated with a particular set of physical properties.\cite{Whetten, Patel, Landman} 

In principle, it should be possible to accuratly describe NCs with a capping layer of so-called ligands and their corresponding superstructures via molecular simulation. Determining the superstructure of NCs from atomistic molecular simulations is, however difficult, because of the large number of interaction sites -- up to several thousand for a single NC -- and most importantly the slow dynamics of the phase transition towards stable superstructures. 

As a consequence, in practice coarse graining stretegies have to be applied. We take the route to first determine the effective pair potential between two NC, the potential of mean force (PMF), from all-atomistic molecular dynamic simulations.\cite{schap1} Calculating phase diagrams of various stable superstructures still becomes tedious, because the PMF for defined NCs (or mixtures of NCs) depends on the surrounding solvent and temperature.

% 
% There are several studies that investigate effects of shape and size of nano crystals on the PMF. In particular the influence of ligand coverage and length is of interest, because interactions between ligands dominate the PMF. [Schapotschnikow] studied the influence of ligand and core sizes and developed a parameterization for the effective potential between two NC. However, the derived parameters are only applicable for the fitted temperatures.    

In this work, we propose a method to predict the PMF between two NCs in vacuum for different temperatures. Using thermodynamic perturbation theory of first order, we develop a correlation of the free energy as a function of the temperature that simplifies the practical application considerably. Parameterizing the perturbation expression with results from Molecular Dynamics (MD) simulations enables the prediction of PMF for a wide range of different temperatures from only two simulated PMF curves.

The method is applied to various systems of gold NC capped with alkyl chains with thiol head groups in vacuum. Variables are the core size and the number and length of attached ligands. The predicted potentials are compared to simulation data and show very good agreement.

The remainder of the paper is structured as follows. First, we describe the simulation setup and summarize the thermodynamic perturbation theory. Then, the results for different systems are presented, followed by a brief conclusion.

%%%%%%%%%%%%METHODS%%%%%%%%%%%%%%% 
\section{Method}
In this section we introduce a method that relates MD simulations to an analytic fluid theory to predict the temperature dependence of the PMF. First, we review how the PMF, representing the effective interactions of capped gold nano crystals at fixed distances, is obtained from molecular simulations. Then we summarize the relevant elements of thermodynamic perturbation theory (TPT) that are needed to model the underlying temperature dependence. For the application of capped NCs, we derive a relation for calculating the PMF at an arbitrary temperature.

\subsection{PMF from molecular dynamic simulations}

% \subsection{PMF from MD simulations}

% The potential of mean force is an effective potential between two tagged molecules in a system containing a large number of molecules. These tagged molecules are fixed in space. The PMF between them contains all the interactions with the surrounding (e.g. solvent molecules). In vacuum, the PMF reduces to the pair potential between the tagged molecules.

There are different methods to obtain the PMF via computer simulations e.g. configurational-bias Monte Carlo, steered Molecular Dynamics or constraint Molecular Dynamics simulations. An overview is given by Trzesniak et al.\cite{Trzesniak}

In this work we use constraint Molecular Dynamics to measure forces acting on the center of mass of each NC in vacuum to be able to calculate the PMF. The NC gold cores are modeled as rigid icosahedra, exposing only (111)-facets.\cite{wangy} According to a united-atom approach, we consider $SH$-, $CH_2$- and $CH_3$-groups in the ligands as single interaction sites with a force field described in the work of Schapotschnikow et al.\cite{schap1} Our simulations are performed in the NVT ensemble using the GROMACS simulation software\cite{Pronk}. In simulations we impose periodic boundary conditions and  the simulation box is sufficiently large to avoid interactions of capping layers of NCs with their periodic images. The length of the simulation runs are 5 ns. For each system 10 runs were conducted for every temperature.

To calculate the PMF we fix two NCs at the desired distance $r_{12}$ and measure the forces acting on the center of mass of each NC, ${\bf F_1}$ and ${\bf F_2}$. The mean force between two particles is then obtained from
\begin{equation*}
      F_{m}(r_{12})=\frac{1}{2}\left\langle\left({\bf F_2}-{\bf F_1}\right)\cdot{\bf r_u} \right\rangle_{NVT},
\end{equation*}
where ${\bf r_u}={\bf r_{12}}/r_{12}$ is the unit vector connecting two particles and the angular brackets denote the average in the canonical ensemble. For every distance $r_{12}$, a separate simulation is needed. In the last step we integrate the obtained 
forces over all distances to get the potential of the mean force, as
\begin{equation*}
  \Phi_{m}(r_{12})=\int\limits_{r_{12}}^\infty \! F_{m}(s) \,\mathrm{d}s.
\end{equation*}

\subsection{Thermodynamic perturbation theory}\label{tpt}

The complex interactions between particles can successfully be modeled via perturbation theory.\cite{Zwanzig, bh1} The starting point is a pair potential between atomistic (or united-atom) interaction sites. This potential is split into two parts. The first part, representing short ranged interactions, is the reference and often chosen to describe repulsive interactions. The second part is refered to as perturbation.

Introducing the coupling paramter $\lambda$, the pair potential reads
\begin{align}
  u_{\lambda} = u^\text{ref} + \lambda u^\text{per}\,, \label{lambda}
\end{align}
where $\lambda\in[0,1]$ switches the perturbation on and off. $u_{\lambda=1}=u$ is the full pair potential while $u_{\lambda=0}=u^\text{ref}$ is the reference.
For any parameter, a Taylor expansion with respect to $\lambda$ is representing the corresponding perturbation. In equation \ref{lambda} and in the following equations we have assumed only one type of (united-atom) interaction site. In the current case it is an average of thiol-, $CH_2$- and $CH_3$-groups of the ligands. It is straight forward to develop all equations for different individual interaction sites, but this is not necessary here. With equation \ref{lambda}, the Helmholtz energy becomes
 \begin{align}
%  A(\rho, T) = A^{HS}(\rho, T) + A^{attr}(\rho, T) 
A^\text{tar} = A^\text{ref} + A^\text{per}\,,
\end{align}
where the perturbation is expanded as
%The total Helmholtz energy of a -- in our case -- dispersive fluid has then two contributions: pure repulsion and attraction.
%The Helmholtz energy is a function of the number of molecules in the system as well as the volume and the temperature. For reasons of simplicity, these dependencies are not written explicitly.
\begin{align}
 A^\text{per} = \left( \frac{\partial A}{\partial \lambda}\right)_{\lambda = 0} \lambda + \frac{1}{2} \left( \frac{\partial² A}{\partial \lambda²}\right)_{\lambda = 0} \lambda²+ \mathcal{O}(\lambda^3) \,,
\end{align}

with $\lambda=1$. In first order,
\begin{align}
 \left( \frac{\partial A}{\partial \lambda}\right)_{\lambda = 0} &= - \frac{1}{\beta} \left(\frac{1}{Z} \frac{\partial }{\partial \lambda} \underbrace{\int \exp(-\beta U_{N,\lambda}(\mathbf{r}^N)) \mathrm{d}\mathbf{r}^N}_{=Z} \right)_{\lambda = 0} \\
 &= \frac{1}{Z^\text{ref}}\int U_N^\text{per}(\mathbf{r}^N) \exp(-\beta U_N^\text{ref}(\mathbf{r}^N)) \mathrm{d}\mathbf{r}^N \,,
\end{align}

where $U_{N,\lambda}(\mathbf{r}^N)$ is the total potential energy of the system and $Z$ denotes the configuration integral and $U_N^\text{per}$ is equivalently the total potential energy of the perturbation part of the potential according to equation \ref{lambda}. With the definition of the pair correlation function $g_{\alpha \beta}^\text{ref}(\mathbf{r}_1,\mathbf{r}_2)$, where $\alpha$ and $\beta$ denote ligand segments of NC1 and NC2 respectively, one gets

% The latter expression yields the definition of the pair distribution function
% \begin{align}
%  g^{ref}(\mathbf{r}_1,\mathbf{r}_2) = \frac{1}{Z^{ref}} \frac{N(N-1)}{\rho^2} \int \exp(-\beta U^{ref}(\mathbf{r}^N))) \mathrm{d}\mathbf{r}^{(N-2)} \,. \label{ref}
% \end{align}
% 
% The first derivative of the Helmholtz energy for a homogeneous system is then
% \begin{align}
%  \left( \frac{\partial A}{\partial \lambda}\right)_{\lambda = 0} = \frac{1}{2}\rho^2 \iint u^{per}(r_{12}) g^{ref}(\mathbf{r}_1,\mathbf{r}_2) \mathrm{d}\mathbf{r}_1 \mathrm{d}\mathbf{r}_2 \,,
% \end{align}
% 
% and the Helmholtz energy ensuing from the pair potential is

\begin{align}
 A = A^\text{ref} +\frac{1}{2} \sum_{\alpha} \sum_{\beta} \iint \rho_{\alpha}(\mathbf{r}_1) \rho_{\beta}(\mathbf{r}_2) g_{\alpha \beta}^\text{ref}(\mathbf{r}_1,\mathbf{r}_2)  u^\text{per}(r_{12}) \mathrm{d}\mathbf{r}_1 \mathrm{d}\mathbf{r}_2 \,. \label{eq:A_tar}
\end{align}

Equation \eqref{eq:A_tar} is the Helmholtz energy (as a functional of the density of ligand segments) according to first-order perturbation theory. Previous studies showed, that ligand interactions dominate effective interactions between NCs.\cite{Tay:2005} Only at small distances, core interactions have to be considered. Therefore, equation \ref{eq:A_tar} contains the ligand segment densities $\rho_{\alpha}$ and $\rho_{\beta}$ only. Ligand interactions can be modeled using a Lennard Jones potential as a target potential. We see that it is crucial to choose a suitable reference, since, to apply perturbation theory, it is necessary to have knowledge of the structure (i.e. $g_{\alpha \beta}^\text{ref}$) of the reference fluid. Therefore, we have chosen a system, where all interaction sites are represented by hard-sphere potentials. The reference fluid then represents a hard-sphere chain (superindex 'hsc') fluid, which is well described by Tripathi and Chapman.\cite{TC} The Helmholtz energy is then

% Since the repulsive partition of the Lennard Jones potential is different from a hard sphere potential, one can not simply substitute the potentials.   
% 
% Depending on the division of the Lennard Jones potential into a reference potential $u^\text{ref}$ and a perturbation potential $u^\text{per}$, we have to modify the hard sphere reference in a way that its contribution to the Helmholtz energy is the same.

\begin{align}
 A = A^\text{hsc} + \frac{1}{2} \sum_{\alpha} \sum_{\beta} \iint \rho_{\alpha}(\mathbf{r}_1) \rho_{\beta}(\mathbf{r}_2) g_{\alpha \beta}^\text{hsc}(\mathbf{r}_1,\mathbf{r}_2)  u^\text{per}(r_{12})\mathrm{d}\mathbf{r}_1 \mathrm{d}\mathbf{r}_2 \,. \label{ghs}
\end{align}

To describe the reference using hard-sphere chains we need to assure that this reference fluid provides the same Helmholtz energy contribution as the reference part of our target potential. One option to achieve this is to modify the hard sphere contact distance by defining an equivalent hard sphere diameter $d(T,\rho)$.\cite{Rowlinson1} It is important to note, that the temperature behavior of $d$  depends on the division of Lennard Jones potential into reference and perturbation. Prominent separations were proposed by Barker and Henderson\cite{bh2} and Weeks, Chandler and Andersen\cite{wca} and are not shown here.

% Following the WCA approach for a Lennard Jones potential, we use the equivalent diameter derived by Verlet and Weis
% \begin{align}
%  d_{VW}(T) = \frac{0.3837T+1,068}{0,4293T+1} \,.
% \end{align}
% [Mulero et al. for range of applicability]. 

We will show later that for our method it is not necessary to actually choose a separation distance explicitly. For a fairly large temperature range, it is sufficient to assume a constant hard-sphere diameter.

% The hard sphere fluid is well described by Carnahan and Starling.\cite{Carnahan} 
We write the dimensionless total Helmholtz energy $A/kT$ as
\begin{align}
 A/kT  = \underbrace{A^\text{hsc}/kT}_{\neq f(T)} + A^\text{per}/kT \,. \label{hs_pert}
\end{align}

In terms of our method, we can simplify this expression and define two functions $a$ and $b$ that depend on the number of molecules and the volume only.
\begin{align}
 A/kT  = a(N,V) + b(N,V)/kT \,, \label{method}
\end{align}

where $b(N,V)$ is a temperature independent correlation given by the last term of equation \ref{ghs}. The temperature dependence through intramolecular potentials is absorbed into an ideal gas contribution which can be assumed to be equal for every center-of-mass distance. Since we are interested in the Helmholtz energy difference, its contribution vanishes.

To summarize we see that using perturbation theory for a Lennard Jones system enables us to formulate a very simple temperature dependence of the Helmholtz energy contributions. This expession assumes a first-order perturbation theory using a constant equivalent hard sphere diameter.

Up to now, we moved along two different paths. The first one yielded the effective potential using atomistic MD simulations, the second one provided us with a functional temperature dependence of the Helmholtz energy for systems with Lennard Jones interactions. To motivate the connection between these two paths, we consider two distances 1 and 2 of two NCs in vacuum, where both configurations are in equilibrium. We then pull the centers of mass towards each other. This way, we transfer configuration 1 into configuration 2. The work that is needed is the potential of mean force and can be formulated as Helmholtz energy difference

\begin{equation}
   \Phi_{m} = A_2 - A_1 \,. \label{eq_jarz}
\end{equation}

This expression is a simple case of Jarzynski's non-equilibrium equality.\cite{Jarzynski, Jarzynski2} It holds under the condition, that the transfer between configurations happens adiabatically slow. In other words, every configuration along the path between 1 and 2 has to be in equilibrium which is exactly what we establish in atomistic MD simulations. We simplify the notation in \ref{eq_jarz}

\begin{equation}
 \Phi_{m}(r_{12}) = \Delta A(r_{12}) \,,
\end{equation}

where $r_{12}$ denotes the separation distance between centers of mass of the two NCs analogue to the previous description. Now, we use the derived formula from perturbation theory \ref{hs_pert} to formulate the right hand side   

\begin{align}
 \Phi_{m}(r_{12})/kT =&  \Delta A^\text{hsc}(r_{12})/kT + \Delta A^\text{per}(r_{12})/kT \\
 =& a(r_{12}) + b(r_{12})/kT \,.
\end{align}

Here, we have made a transition from the variables $N,V$ of equation \ref{method} to the separation distance of two NCs, $r_{12}$. That is possible, because in equilibrium conditions, a given $r_{12}$ uniquely determines the average density field and thus $\langle N \rangle$ for a defined $V$.

$a(r_{12})$ and $b(r_{12})$ can be calculated directly from simulations at two different temperatures

\begin{align}
  a(r_{12}) &= \frac{\Phi_{m}(r_{12},T_2)-\Phi_{m}(r_{12},T_1)}{k(T_2-T_1)} \\
  b(r_{12}) &= \frac{T_1 \Phi_{m}(r_{12},T_2)-T_2 \Phi_{m}(r_{12},T_1)}{T_1-T_2} \,.
\end{align}

The elegance of this method is that it is not necessary to concern oneself with the actual decomposition of the target potential or the calculation of the Helmholtz energy contributions while still beeing able able to utilize the theoretical framework from perturbation theory to reduce the simulation effort drastically.

%%%%%%%%%%%%%%%%RESULTS%%%%%%%%%%%%%%
\section{Results}

We investigated NCs consisting of 147 and 1415 gold atoms that form the core. Attached are alkyl thiols with 8 to 12 carbon atoms. For all systems the cores are fully loaded with ligands.\cite{Schap2} For each PMF, the initial separation distance was chosen sufficiently large to assure no significant interactions between the two NCs. In this region the PMF is zero. The simulation results were verified in comparison to the results obtained by Schapotschnikow et al.\cite{schap1} 

\begin{figure}[t!]
  \begin{center}
    \subfloat[$Au_{147}(SC8)_{58}$]{\includegraphics[width=0.48\textwidth]{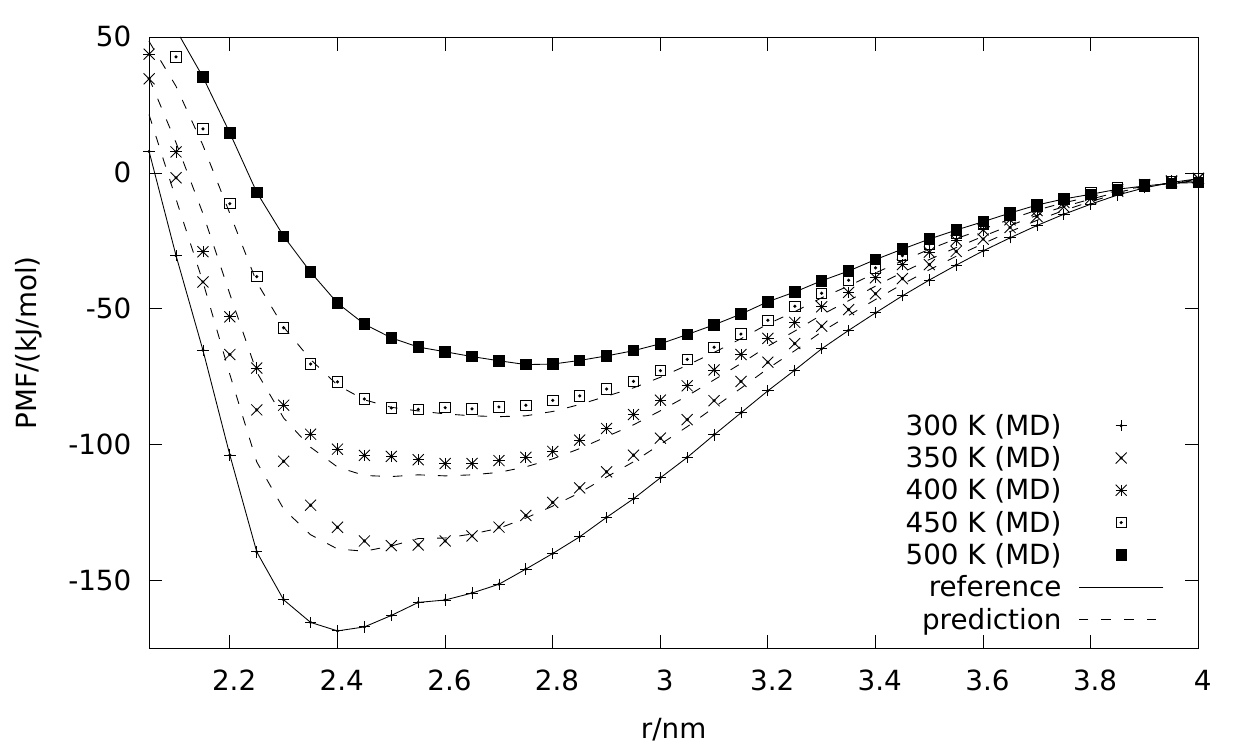}\label{Au147_8}}
    \quad
    \subfloat[$Au_{147}(SC12)_{58}$]{\includegraphics[width=0.48\textwidth]{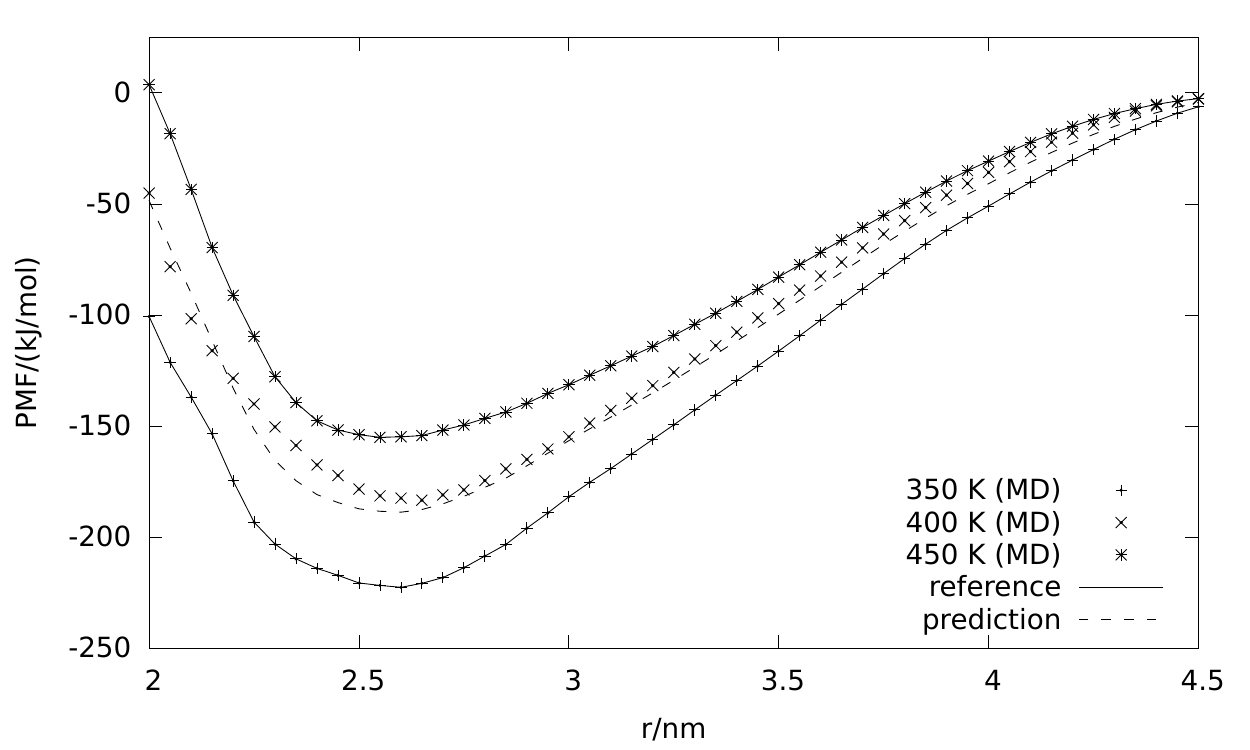}    \label{Au147_12}}
  \caption{PMF as function of center of mass separation $r$ between two nano crystals for different temperatures in vacuum. Symbols represent data from constraint MD simulations, solid lines represent reference PMF and dashed lines denote predicted PMF with the presented method.}
  \label{Au147}
  \end{center}
\end{figure}

Figure \ref{Au147} presents the PMF from simulations for a system of two NCs consisting of 147 gold atoms with 58 ligands. The ligands consist of alkyl chains with 8 (Figure \ref{Au147_8}) and 12 (Figure \ref{Au147_12}) carbon atoms (notation $Au_{147}(SC8)_{58}$ and $Au_{147}(SC12)_{58}$). The diagrams present the PMF for varying center-of-mass distances for different temperatures. Every symbol denotes a total number of 10 simulations. Dashed lines represent predicted potentials using the method presented in section \ref{tpt}. Solid lines represent the PMF used as reference.  

The predicted PMF show very good agreement in the whole region of the simulations. Even though there are minor deviations from simulation data near the minima, all predicted values lie within error bars. Most important, the attractive region, where error bars from simulations are small, is predicted with good agreement. It is noteworthy, that the references should be chosen at temperatures where ligands have no prefered orientations which occurs in melting and freezing transitions.\cite{Landman2, Badia}     

\begin{figure}[t!]
  \begin{center}
    \subfloat[$Au_{1415}(SC12)_{242}$]{\includegraphics[width=0.48\textwidth]{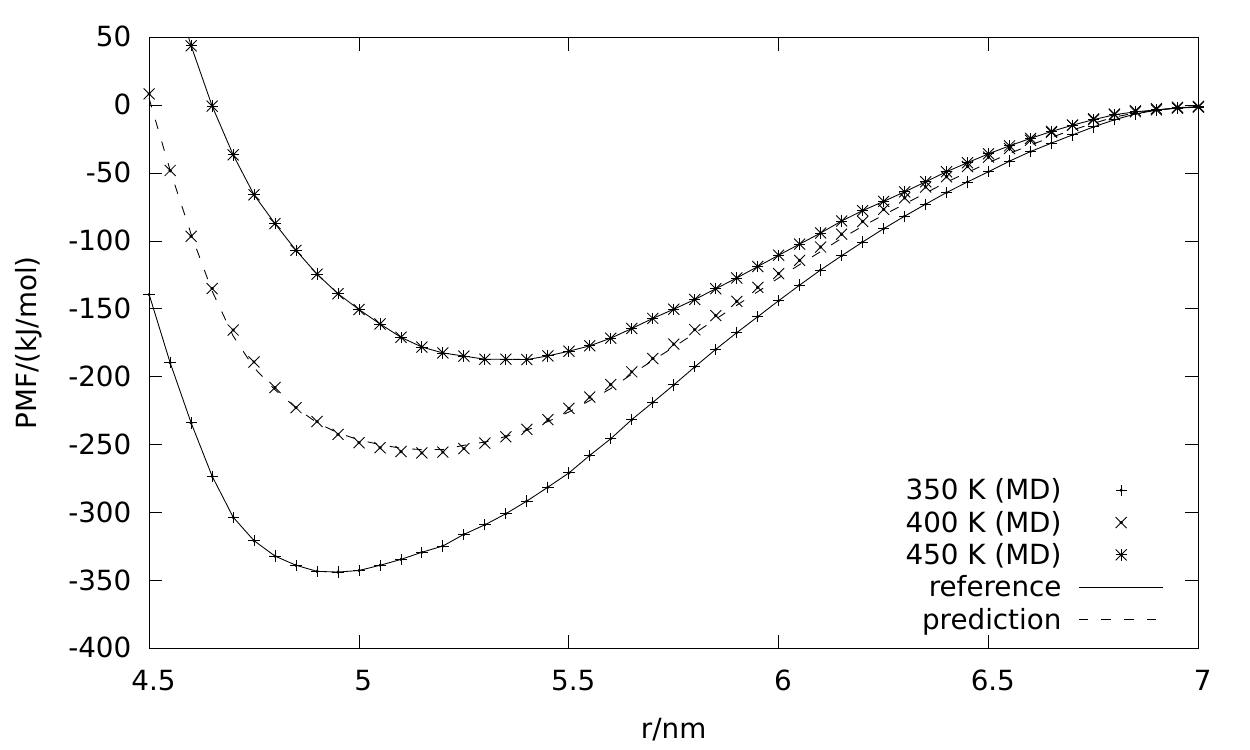}\label{Au1415_12}}
    \quad
    \subfloat[Decomposition of the Helmholtz energy into reference and perturbation contribution.]{\includegraphics[width=0.48\textwidth]{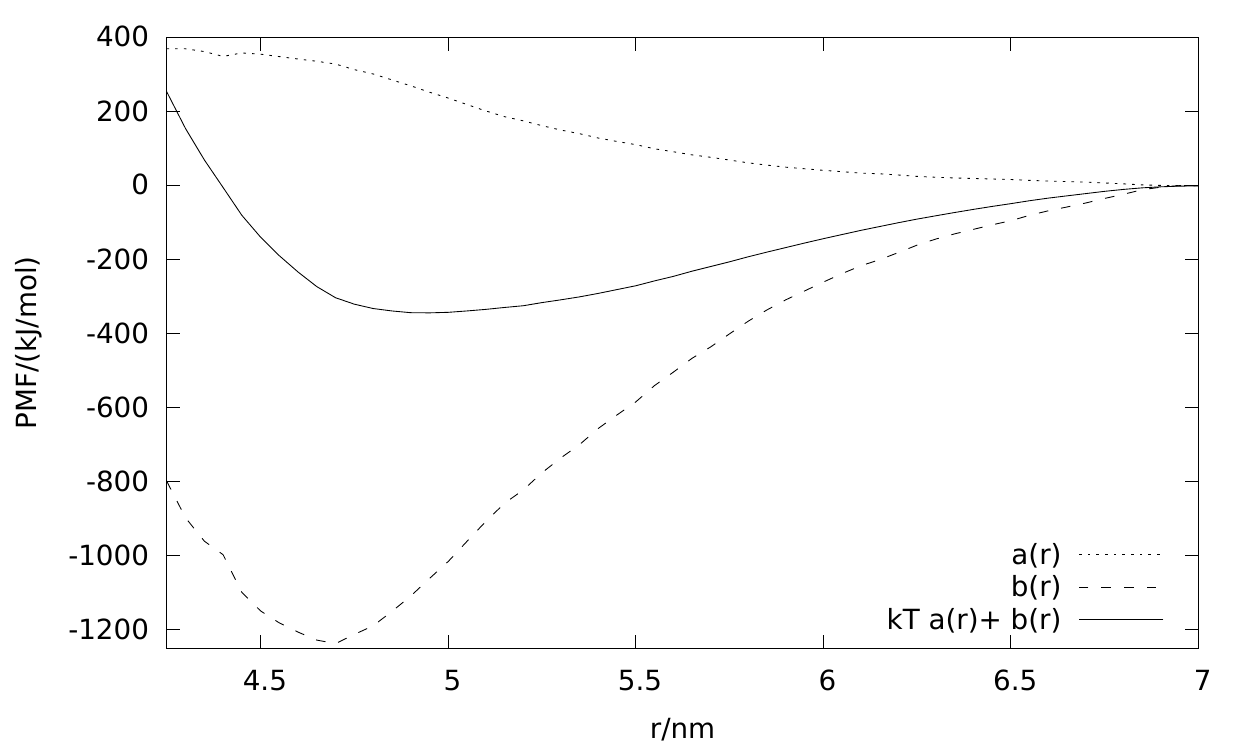}    \label{ref_pert}}
  \caption{PMF as function of center of mass separation $r$ between two nano crystals for different temperatures in vacuum. The reference and perturbation contribution are calculated using the PMF data at 350~K and 450~K.}
  \label{Au1415}
  \end{center}
\end{figure}

Figure \ref{Au1415} shows results for a system of two NCs consisting of 1415 gold atoms and 242 ligands with 12 carbon atoms. Figure \ref{Au1415_12} presents the predicted PMF while Figure \ref{ref_pert} illustrates profiles for the functions $a(r_{12})$ and $b(r_{12})$ and the predicted PMF.

%%%%%%%%%%%%%CONCLUSION%%%%%%%%%%%%%%
\section{Conclusion}
In this article we propose a method to predict the PMF between capped gold NCs in vacuum at any temperature using data from only two MD simulations. The underlying dependency of the PMF on the temperature was derived from first-order perturbation theory where we simplified the reference to be only a linear function of the temperature. The predicted PMF were compared to results from constraint MD simulations and showed very good agreement for systems of different sized and shaped NCs.

\bibliography{literature}
\bibliographystyle{abbrv}
\end{document}